\documentclass[aip,jmp,preprint,numerical]{revtex4-1}

\usepackage{amsmath}
\usepackage{amssymb}
\usepackage{amsfonts}
\usepackage{graphicx}
\usepackage{dcolumn}
\usepackage{bm}
\usepackage{hyperref}
\hypersetup{colorlinks=false}

\usepackage[dvips]{color}

\begin{document}

\title[Some exact results on quantum Newtonian cosmology]{Some exact results on quantum Newtonian cosmology}

\date{\today}

\author{H. S. Vieira}
\email{horacio.santana.vieira@hotmail.com}
\affiliation{Departamento de F\'{i}sica, Universidade Federal da Para\'{i}ba, Caixa Postal 5008, CEP 58051-970, Jo\~{a}o Pessoa, PB, Brazil}
\author{V. B. Bezerra}
\email{valdir@fisica.ufpb.br}
\affiliation{Departamento de F\'{i}sica, Universidade Federal da Para\'{i}ba, Caixa Postal 5008, CEP 58051-970, Jo\~{a}o Pessoa, PB, Brazil}
\author{C. R. Muniz}
\email{celio.muniz@uece.br}
\affiliation{Grupo de F\'isica Te\'orica (GFT), Universidade Estadual do Cear\'a, Faculdade de Educa\c c\~ao, Ci\^encias e Letras de Iguatu, Iguatu, Cear\'a, Brazil}
\author{M. S. Cunha}
\email{marcony.cunha@uece.br}
\affiliation{Grupo de F\'isica Te\'orica (GFT), Centro de Ci\^encias e Tecnologia, Universidade Estadual do Cear\'a, CEP 60714-903, Fortaleza, Cear\'a, Brazil}

\begin{abstract}
We obtain the wave functions associated to the quantum Newtonian universe with a cosmological constant which is described by the Schr\"{o}dinger equation and discuss some aspects of its dynamics for all forms of energy density, namely, matter, radiation, vacuum, dark energy, and quintessence. These wave functions of the quantum Newtonian universe are obtained in terms of the Heun's functions and the respective energy levels are shown. We use these solutions to investigate the expansion of the universe and found that the asymptotic behavior for the scale factor is $R \sim \mbox{e}^{t}$ for whatever the form of energy density is. We also analyze the behavior of the universe at early stages.
\end{abstract}




\maketitle


%
%
\section{Introduction}
A model of the universe constructed by combining Newtonian mechanics with the cosmological principle and assuming the fact that the matter is pressureless and our universe experiences an expansion was presented in the 1930's by Milne and McCrea \cite{QJMath.5.64,QJMath.5.73}. Then, they derived a cosmological equation which is algebraically analogous to the Friedmann one, to describe the evolution of the universe taking into account the expression for the total energy of the particles (galaxies) subject to the gravitational interaction and additionally, considering the cosmological principle. Taking into account this system, we can construct the Lagrangian and write down the Hamiltonian associated to it and find the equation and integral of motion. The description obtained in this approach, with pressureless matter, is equivalent to the one obtained in the context of relativistic cosmology. In other words, the description of the universe using Newtonian dynamics and gravitation, for a pressureless system and assuming the cosmological principle and the expansion of the universe, is exactly the same obtained in the context of relativistic cosmology.

In what follows we will use this classical approach to describe the universe using Newtonian mechanics and gravitation and extend it to the domain of quantum cosmology. To do this we construct the Hamiltonian of the system from which we can write the corresponding Schr\"{o}dinger equation whose wave function is supposed to describe the time-dependent evolution of the universe, which should depend only on the scale factor. Thus, this wave function describes the evolution of the observed universe and this evolution follows a unique route. Therefore, using this point of view, it is not necessary to take into account the Schr\"{o}dinger equation for a system of $N$ particles, with $N > 1$, but just consider the single particle problem.

It is worth calling attention to the fact that when the pressure is different from zero, it is necessary to modify the equations to take into account the pressure \cite{ProcRSocLondA.206.562} and adopt a series of assumptions \cite{ProcRSocLondA.149.384} to guarantee that the Newtonian and Einstein theories give similar results.

In the context of pure quantum cosmology the universe should be described by a single wave function defined for the different Friedmann-Robertson-Walker spacetime geometries taking into account all matter contents. In this approach, the equation for the wave function is not the Schr\"{o}dinger equation, but the Wheeler-DeWitt equation \cite{PhysRev.160.1113,Wheeler:1968}. Based on this equation a lot of investigations have been done with the proposal to determine the wave function of the universe \cite{PhysLettB.117.25,PhysRevD.27.2848,PhysRevD.28.2960,LettNuovoCimento.39.401,PhysRevD.30.509,NuclPhysB.239.257,NuclPhysB.252.141,PhysRevD.32.2511,NuclPhysB.264.185,PhysRevLett.57.2244,PhysLettA.236.10,PhysRevD.86.063504,ClassQuantumGrav.30.143001}.

On the other hand, it is possible to find the wave function of the universe in the context which corresponds to what is termed quantum Newtonian cosmology. In this approach the wave functions are solutions of the Schr\"{o}dinger equation for the system under consideration \cite{AIPConfProc.743.286,arXiv:0504072,ProcRSoc.A.463.503,IntJTheorPhys.47.455,ISRNMathPhys.2013.509316}, which, in principle, is much more simple than in pure quantum cosmology approach where the Wheeler-DeWitt equation is taken into account. Otherwise, it is more involved than by merely adding quantum corrections to the Newtonian potential \cite{EurPhysJC.76.543}.

Thus, we will consider different epochs of the early Universe, where the involved energies are lower but its scale is small enough in order to use a quantum approach by taking the Schr\"{o}dinger equation into consideration. In this scenario, the galaxies do not have emerged yet, but their seeds, yes. This quantum approach to Newtonian cosmology does not correspond necessarily to a real description of the evolution of the universe, but could be a source of inspiration to construct, possibly, a real quantum theory of gravity.

It is worth calling attention to the fact that in adopting this point of view, we are extending the Birkhoff's theorem to quantum Newtonian cosmology. We considered the usual understanding of standard cosmology, in which a galaxy is a material point under the influence of $1/R$ potential generated by a spherical distribution of matter with radius $R$ reaching it. The galaxies outside this region are not taken into account. In addition, we consider later epochs of the universe, where the energies are lower but the scale factor is small enough in order that it is possible to use of the Schr\"{o}dinger equation to describe the system, instead of the Wheeler-DeWitt equation.

This paper is organized as follows. In Section~\ref{Sec.II}, we solve the Schr\"{o}dinger equation for different contents of matter, show the effective potential energy and determine the energy spectrum. In Section~\ref{Sec.III}, we analyze the behaviors of the scale  factor for different scenarios. Finally, in Section~\ref{Sec.IV}, we present the conclusions.
\newpage
%
%
\section{Schr\"{o}dinger equation in a Newtonian universe: wave functions and energy levels}\label{Sec.II}
\begin{figure}[t]
		\includegraphics[scale=0.50]{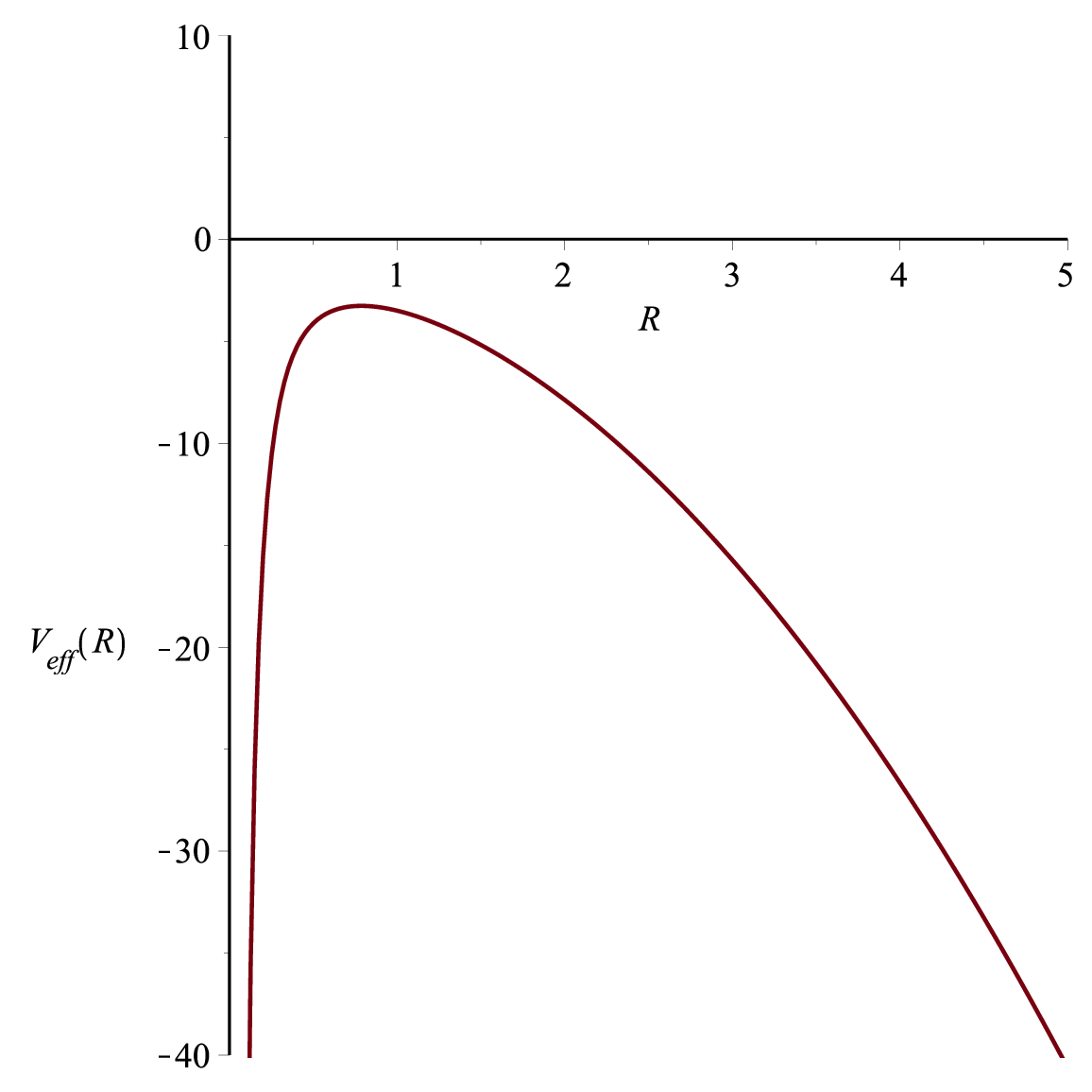}
	\caption{The effective potential energy for $\Lambda > 0$.}
	\label{fig:Newtonian_universe_Fig1}
\end{figure}

\begin{figure}[t]
		\includegraphics[scale=0.50]{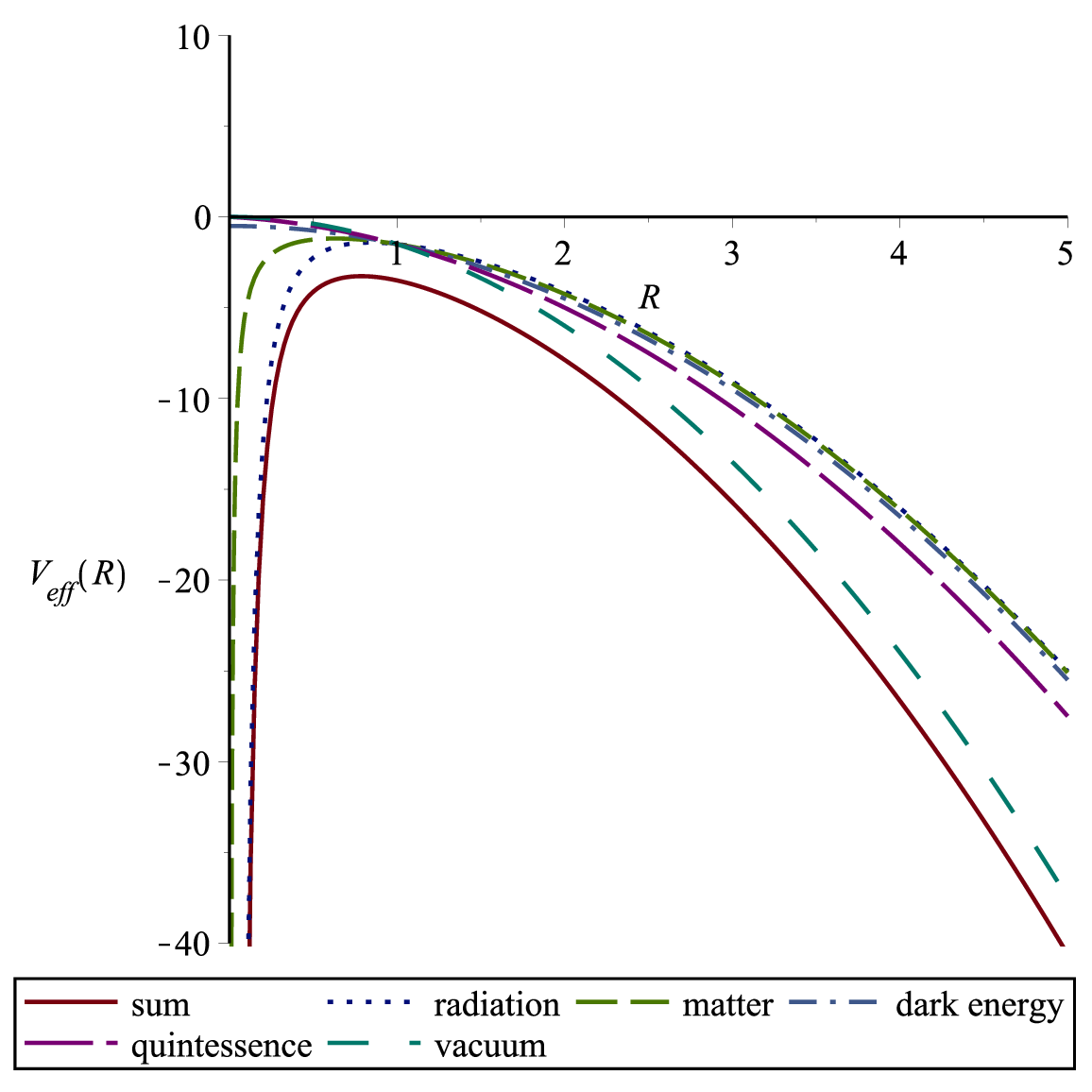}
	\caption{The effective potential energy for $\Lambda > 0$, for different kind of energy densities.}
	\label{fig:Newtonian_universe_Fig2}
\end{figure}

\begin{figure}[t]
		\includegraphics[scale=0.50]{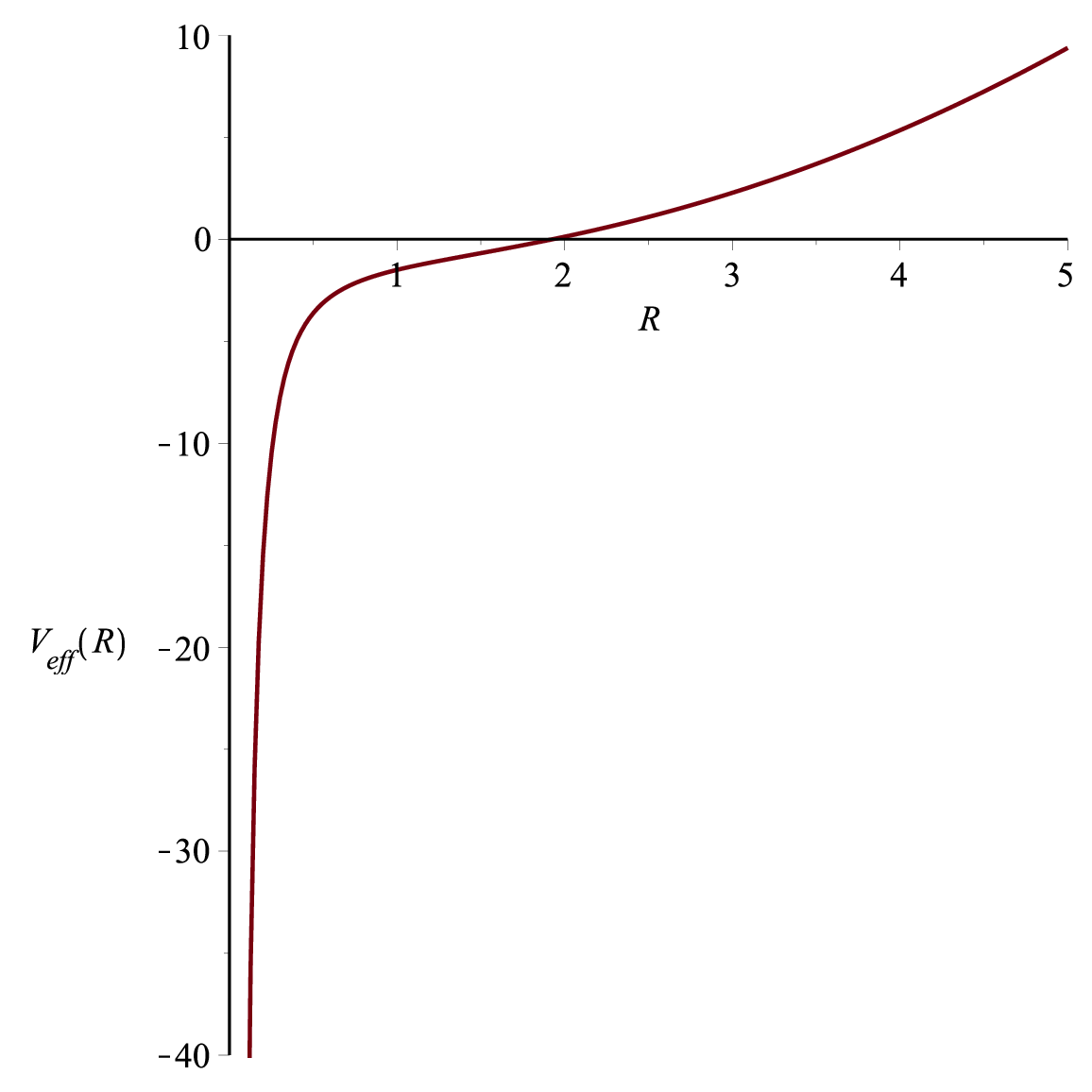}
	\caption{The effective potential energy for $\Lambda = -|\Lambda|$.}
	\label{fig:Newtonian_universe_Fig3}
\end{figure}

\begin{figure}[t]
		\includegraphics[scale=0.50]{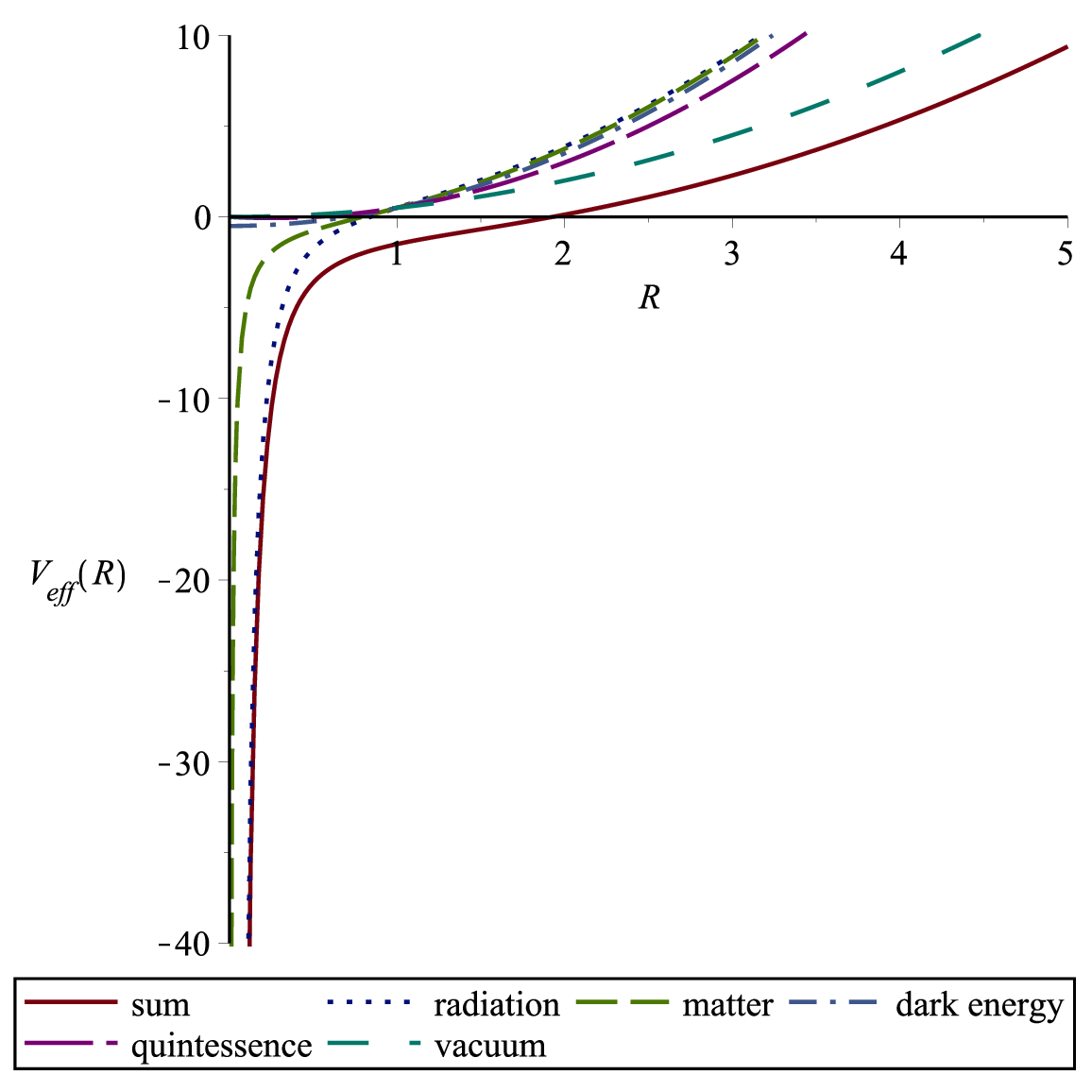}
	\caption{The effective potential energy for $\Lambda = -|\Lambda|$, for different kind of energy densities.}
	\label{fig:Newtonian_universe_Fig4}
\end{figure}

In a previous paper \cite{JMathPhys.56.092501}, we have found that the Hamiltonian operator for a particle (in what follows when we refer to particle, means galaxy) moving in the Newtonian universe is given by
\begin{equation}
H=-\frac{\hbar^{2}}{2\mu}\frac{d^{2}}{dR^{2}}-\frac{GM\mu}{R}-\frac{1}{6}\Lambda \mu R^{2}\ ,
\label{eq:Hamiltonian_op-canonical}
\end{equation}
where $\mu$ is the mass of a particle, $R$ is the scale factor, and $\Lambda$ is the cosmological constant. The total mass $M$ of the Newtonian universe (mass inside the sphere of radius $R$) is given by
\begin{equation}
M=\frac{4}{3}\pi R^{3}\rho\ .
\label{eq:mass_Newtonian_universe}
\end{equation}
The density energy can be expressed as \cite{PhysRevD.94.023511}
\begin{equation}
\rho_{\omega}=A_{\omega}R^{-3(\omega+1)}\ ,
\label{eq:WDE_density}
\end{equation}
where
\begin{equation}
A_{\omega}=\rho_{\omega 0}R_{0}^{3(\omega+1)}\ ,
\label{eq:A_WDE_density}
\end{equation}
and $\rho_{\omega 0}$ stands for the value of $\rho_{\omega}$ at present time, with the state parameter, $\omega$, being given by
\begin{equation}
\omega=\left\{
\begin{array}{rl}
	0            & \mbox{for matter}\ (\rho_{m})\ ,\\
	\frac{1}{3}  & \mbox{for radiation}\ (\rho_{r})\ ,\\
	-1           & \mbox{for vacuum}\ (\rho_{v})\ ,\\
	-\frac{1}{3} & \mbox{for dark energy}\ (\rho_{d})\ ,\\
	-\frac{2}{3} & \mbox{for quintessence}\ (\rho_{q})\ ,
\end{array}
\right.
\label{eq:omega_Newtonian_universe}
\end{equation}
so that
\begin{equation}
\left\{
\begin{array}{l}
	A_{m}=\rho_{m0}R_{0}^{3}\ ,\\
	A_{r}=\rho_{r0}R_{0}^{4}\ ,\\
	A_{v}=\rho_{v0}\ ,\\
	A_{d}=\rho_{d0}R_{0}^{2}\ ,\\
	A_{q}=\rho_{q0}R_{0}\ .
\end{array}
\right.
\label{eq:A_Newtonian_universe}
\end{equation}
If matter, radiation, vacuum, dark energy and quintessence contribute all together, the total energy density $\rho$ is given by the sum
\begin{equation}
\rho=\sum_{\omega}\rho_{\omega}=\rho_{m}+\rho_{r}+\rho_{v}+\rho_{d}+\rho_{q}\ .
\label{eq:density_sums}
\end{equation}
Thus, substituting Eqs.~(\ref{eq:mass_Newtonian_universe})-(\ref{eq:omega_Newtonian_universe}) into Eq.~(\ref{eq:Hamiltonian_op-canonical}), we obtain
\begin{equation}
H=-\frac{\hbar^{2}}{2\mu}\frac{d^{2}}{dR^{2}}+V_{eff}(R)\ ,
\label{eq:generalized_Hamiltonian_Newtonian_universe}
\end{equation}
which corresponds to the generalized Hamiltonian operator for a particle moving in the Newtonian universe, where $V_{eff}(R)$ is the effective potential energy given by
\begin{equation}
V_{eff}(R)=-\frac{4 \pi G \mu}{3}\biggl[A_{d}+A_{q}R+\biggl(A_{v}+\frac{\Lambda}{8 \pi G}\biggr)R^{2}+\frac{A_{m}}{R}+\frac{A_{r}}{R^{2}}\biggr]\ .
\label{eq:Newtonian_universe_effective_potential_energy}
\end{equation}
The behaviors of $V_{eff}(R)$ for all cases are shown in Figs.~\ref{fig:Newtonian_universe_Fig1}-\ref{fig:Newtonian_universe_Fig4}, for positive and negative values of the cosmological constant.

In the Newtonian approach to cosmology, the dynamics of the system comes from the radial proper motion of the galaxies, and not of the spacetime itself, which is supposed to be flat. In this context, the cosmological equation analogous to the Friedmann one is obtained. Thus, let us solve the time independent Schr\"{o}dinger equation $H\psi(R)=E\psi(R)$, where $\Psi(R,t)=\psi(R)\mbox{e}^{-iEt/\hbar}$, and the Hamiltonian is given by Eq.~(\ref{eq:generalized_Hamiltonian_Newtonian_universe}). Instead of solve this equation separately for each value of the state parameter $\omega$, we will do this in a general way, and thus it will be valid for any kind of energy $\omega$. Therefore, the Schr\"{o}dinger equation for a particle moving in the Newtonian universe can be written as
\begin{equation}
\frac{d^{2}\psi(R)}{dR^{2}}+\biggl(B_{1}+B_{2}R+B_{3}R^{2}+\frac{B_{4}}{R}+\frac{B_{5}}{R^{2}}\biggr)\psi(R)=0\ ,
\label{eq:generalized_Schrodinger_equation_Newtonian_universe}
\end{equation}
where the coefficients $B_{1}$, $B_{2}$, $B_{3}$, $B_{4}$, and $B_{5}$ are given by
\begin{equation}
B_{1}=\frac{2\mu E}{\hbar^{2}}+\frac{8 \pi G \mu^{2}}{3 \hbar^{2}}A_{d}\ ,
\label{eq:B1_Newtonian_universe}
\end{equation}
\begin{equation}
B_{2}=\frac{8 \pi G \mu^{2}}{3\hbar^{2}}A_{q}\ ,
\label{eq:B2_Newtonian_universe}
\end{equation}
\begin{equation}
B_{3}=\frac{8 \pi G \mu^{2}}{3 \hbar^{2}}\biggl(A_{v}+\frac{\Lambda}{8 \pi G}\biggr)\ ,
\label{eq:B3_Newtonian_universe}
\end{equation}
\begin{equation}
B_{4}=\frac{8 \pi G \mu^{2}}{3 \hbar^{2}}A_{m}\ ,
\label{eq:B4_Newtonian_universe}
\end{equation}
\begin{equation}
B_{5}=\frac{8 \pi G \mu^{2}}{3 \hbar^{2}}A_{r}\ .
\label{eq:B5_Newtonian_universe}
\end{equation}
Now, we define a new variable, $x$, such that
\begin{equation}
x=\tau R\ ,
\label{eq:x_Newtonian_universe}
\end{equation}
where the parameter $\tau$ is given by
\begin{equation}
\tau=(-B_{3})^{\frac{1}{4}}\ .
\label{eq:tau_Newtonian_universe}
\end{equation}
Thus, with this new variable, we can write Eq.~(\ref{eq:generalized_Schrodinger_equation_Newtonian_universe}) as
\begin{equation}
\frac{d^{2}\psi(x)}{dx^{2}}+\biggl(b_{1}+b_{2}x-x^{2}+\frac{b_{4}}{x}+\frac{b_{5}}{x^{2}}\biggr)\psi(x)=0\ ,
\label{eq:generalized_motion_Newtonian_universe}
\end{equation}
where the coefficients $b_{1}$, $b_{2}$, $b_{4}$, and $b_{5}$ are written as
\begin{equation}
b_{1}=\frac{B_{1}}{\tau^{2}}\ ,
\label{eq:b1_Newtonian_universe}
\end{equation}
\begin{equation}
b_{2}=\frac{B_{2}}{\tau^{3}}\ ,
\label{eq:b2_Newtonian_universe}
\end{equation}
\begin{equation}
b_{4}=\frac{B_{4}}{\tau}\ ,
\label{eq:b4_Newtonian_universe}
\end{equation}
\begin{equation}
b_{5}=B_{5}\ .
\label{eq:b5_Newtonian_universe}
\end{equation}

In what follows we will solve Eq.~(\ref{eq:generalized_motion_Newtonian_universe}) and determine the eigenvalues. Equation (\ref{eq:generalized_motion_Newtonian_universe}) is a biconfluent Heun equation \cite{Ronveaux:1995}, which is a particular case of a second order linear differential equation with four singularities, called Heun equation. The confluent form of this equation is obtained when two of the singularities coalesce and at infinity there is an irregular point. The canonical form of the biconfluent Heun equation reads as
\begin{equation}
\frac{d^{2}y(x)}{dx^{2}}+\biggl(\frac{1+\alpha}{x}-\beta-2x\biggr)\frac{dy(x)}{dx}+\biggl\{(\gamma-\alpha-2)-\frac{1}{2}[\delta+(1+\alpha)\beta]\frac{1}{x}\biggr\}y(x)=0\ ,
\label{eq:Biconfluent_Heun_Canonical}
\end{equation}
where $y(x)=\mbox{HeunB}(\alpha,\beta,\gamma,\delta;x)$ is the biconfluent Heun function. By using the approach described in \cite{AnnPhys.350.14}, we can write Eq.~(\ref{eq:Biconfluent_Heun_Canonical}) in the normal form as
\begin{equation}
\frac{d^{2}Y(x)}{dx^{2}}+\biggl[\frac{1}{4}(4\gamma-\beta^{2})-\beta x-x^{2}-\frac{\delta/2}{x}-\frac{(\alpha^{2}-1)/4}{x^{2}}\biggr]Y(x)=0\ ,
\label{eq:Biconfluent_Heun_normal}
\end{equation}
where $Y(x)=x^{\frac{1}{2}(1+\alpha)}\mbox{e}^{-\frac{1}{2}(x^{2}+\beta x)}y(x)$.

Thus, the Schr\"{o}dinger equation for a particle moving in the Newtonian universe, given by Eq.~(\ref{eq:generalized_motion_Newtonian_universe}) for an arbitrary $\omega$, is similar to the biconfluent Heun function given by Eq.~(\ref{eq:Biconfluent_Heun_normal}), and therefore, its exact solution is given by
\begin{equation}
\psi(x)=C_{1}\ x^{\frac{1}{2}(1+\alpha)}\ \mbox{e}^{-\frac{1}{2}(x^{2}+\beta x)}\ \mbox{HeunB}(\alpha,\beta,\gamma,\delta;x)\ ,
\label{eq:psi_HeunB_Newtonian_universe}
\end{equation}
where $C_{1}$ is a constant to be determined, and the parameters $\alpha$, $\beta$, $\gamma$, and $\delta$ are identified as
\begin{equation}
\alpha=\sqrt{1-4b_{5}}\ ,
\label{eq:alpha_Newtonian_universe}
\end{equation}
\begin{equation}
\beta=-b_{2}\ ,
\label{eq:beta_Newtonian_universe}
\end{equation}
\begin{equation}
\gamma=b_{1}+\frac{b_{2}^{2}}{4}\ ,
\label{eq:gamma_Newtonian_universe}
\end{equation}
\begin{equation}
\delta=-2b_{4}\ .
\label{eq:delta_Newtonian_universe}
\end{equation}
The complete set of solutions of Eq.~(\ref{eq:generalized_motion_Newtonian_universe}), for different values of $\omega$, is summarized in Table \ref{tab:parameters}.

\begin{table}
	\caption{The parameters $\alpha$, $\beta$, $\gamma$, and $\delta$ for the biconfluent Heun function related to the $\omega$ predominance, where $y(x)=\mbox{HeunB}(\alpha,\beta,\gamma,\delta;x)$.}
\label{tab:parameters}
		\begin{ruledtabular}
			\begin{tabular}{cccccc}
			$\omega$ & $\psi(x)$ & $\alpha$ & $\beta$ & $\gamma$ & $\delta$ \\
			\hline
			0 & $x\ \mbox{e}^{-\frac{x^{2}}{2}}\ y(x)$ & 1 & 0 & $\frac{2 \mu E}{\hbar^{2}\tau^{2}}$ & -$\frac{16 \pi G \mu^{2} A_{m}}{3 \hbar^{2} \tau}$ \\
			$\frac{1}{3}$ & $x^{\frac{1}{2}(1+\alpha)}\ \mbox{e}^{-\frac{x^{2}}{2}}\ y(x)$ & $\sqrt{1-\frac{32 \pi G \mu^{2}A_{r}}{3\hbar^{2}}}$ & 0 & $\frac{2 \mu E}{\hbar^{2}\tau^{2}}$ & 0 \\
			-1 & $x\ \mbox{e}^{-\frac{x^{2}}{2}}\ y(x)$ & 1 & 0 & $\frac{2 \mu E}{\hbar^{2}\tau^{2}}$ & 0 \\
			-$\frac{1}{3}$ & $x\ \mbox{e}^{-\frac{x^{2}}{2}}\ y(x)$ & 1 & 0 & $\frac{2 \mu E}{\hbar^{2}\tau^{2}}+\frac{8 \pi G \mu^{2}A_{d}}{3\hbar^{2}\tau^{2}}$ & 0 \\
			-$\frac{2}{3}$ & $x\ \mbox{e}^{-\frac{1}{2}(x^{2}+\beta x)}\ y(x)$ & 1 & -$\frac{8 \pi G \mu^{2}A_{q}}{3\hbar^{2}\tau^{3}}$ & $\frac{2 \mu E}{\hbar^{2}\tau^{2}}+\frac{1}{4}\bigl(\frac{8 \pi G \mu^{2}A_{q}}{3\hbar^{2}\tau^{3}}\bigr)^{2}$ & 0 \\
			\end{tabular}
		\end{ruledtabular}
\end{table}

%
%
As it was described in the paper \cite{JMathPhys.56.092501}, the biconfluent Heun function becomes a polynomial of degree $n$ if and only if the following conditions are fulfilled: $\gamma-\alpha-2=2n$ and $C_{n+1}=0$, where $n=0,1,2,\ldots$, and $C_{n+1}$ is a polynomial in $\delta$. From the first condition, the energy levels for each value of $\omega$ are shown in Table \ref{tab:energy_levels}.

\begin{table}
	\caption{The energy levels related to the $\omega$ predominance.}
	\label{tab:energy_levels}
		\begin{ruledtabular}
			\begin{tabular}{ccccccccc}
			\multicolumn{5}{c}{$\omega$}       & \multicolumn{3}{c}{$E_{n}$} \\
			\hline
			& & 0 & & & & $\frac{\hbar^{2}\tau^{2}}{\mu}\bigl(n+\frac{3}{2}\bigr)$ & \\
			& & $\frac{1}{3}$  & & & & $\frac{\hbar^{2}\tau^{2}}{\mu}\bigl(n+1+\frac{1}{2}\sqrt{1-\frac{32 \pi G \mu^{2}A_{r}}{3\hbar^{2}}}\bigr)$ & \\
			& & -1             & & & & $\frac{\hbar^{2}\tau^{2}}{\mu}\bigl(n+\frac{3}{2}\bigr)$ & \\
			& & -$\frac{1}{3}$ & & & & $\frac{\hbar^{2}\tau^{2}}{\mu}\bigl(n+\frac{3}{2}-\frac{4 \pi G \mu^{2}A_{d}}{3\hbar^{2}\tau^{2}}\bigr)$ & \\
			& & -$\frac{2}{3}$ & & & & $\frac{\hbar^{2}\tau^{2}}{\mu}\bigl[n+\frac{3}{2}-\frac{1}{8}\bigl(\frac{8 \pi G \mu^{2}A_{q}}{3\hbar^{2}\tau^{3}}\bigr)^{2}\bigr]$ & \\
			\end{tabular}
		\end{ruledtabular}
\end{table}

%
%
Now, we adapt the method developed by He \textit{et al}. \cite{PhysLettB.748.361}, based on the approach used by Vilenkin \cite{PhysRevD.33.3560}, who analyzed the dynamical interpretation of the wave function of the universe, from the Wheeler-DeWitt equation in the minisuperspace model. Thus, we will use the exact solution of the Schr\"{o}dinger equation for a particle moving in the Newtonian universe in order to study the dynamical interpretation of the quantum Newtonian wave function and then discuss the boundary conditions in quantum Newtonian cosmology scenario.

As in Eq.~(\ref{eq:generalized_Schrodinger_equation_Newtonian_universe}) there is only one variable, namely, the scale factor $R$, and thus the function $\psi(R)$ can be rewritten as
\begin{equation}
\psi(R)=F(R)\ \mbox{e}^{iS(R)}\ ,
\label{eq:psi_expansion}
\end{equation}
where $F$ e $S$ are real functions. In our case, the square modulus of the wave function of the Newtonian universe is given by
\begin{equation}
|\psi(R)|^{2}=F^{2}(R)\ .
\label{eq:square_modulus_Newtonian_universe}
\end{equation}

From the quantum mechanics formalism \cite{Bransden:2000}, we have that the conserved probability current density is written as
\begin{equation}
\vec{j}(R,t)=\frac{i\hbar}{2\mu}[\Psi^{*}(\vec{\nabla}\Psi)-\Psi(\vec{\nabla}\Psi^{*})]\ ,
\label{eq:current_density_Newtonian_universe}
\end{equation}
in such a way to garantee the validity of the continuity equation, namely,
\begin{equation}
\vec{\nabla} \cdot \vec{j}(R,t)=0\ .
\label{eq:conserved_Newtonian_universe}
\end{equation}
Substituting Eq.~(\ref{eq:psi_expansion}) into Eq.~(\ref{eq:current_density_Newtonian_universe}), we obtain
\begin{equation}
\vec{j}=-\frac{\hbar}{\mu}F^{2}\frac{\partial S}{\partial R}\ .
\label{eq:j_1_Newtonian_universe}
\end{equation}
On the other hand, integrating Eq.~(\ref{eq:conserved_Newtonian_universe}), we get
\begin{equation}
\vec{j}=C_{0}\ ,
\label{eq:j_2_Newtonian_universe}
\end{equation}
where $C_{0}$ is a constant. Thus, from Eqs.~(\ref{eq:j_1_Newtonian_universe}) and (\ref{eq:j_2_Newtonian_universe}), we get
\begin{equation}
-\frac{\hbar}{\mu}F^{2}\frac{\partial S}{\partial R}=C_{0}\ .
\label{eq:result_1_Newtonian_universe}
\end{equation}

Now, we may use the Hamilton-Jacobi formalism of quantum mechanics in order to write down the following relation between the action and the canonical momentum
\begin{equation}
p_{R}=\frac{\partial S}{\partial R}=\frac{\partial L}{\partial \dot{R}}=\mu\dot{R}\ ,
\label{eq:action_momentum_Newtonian_universe}
\end{equation}
where $L$ is the Lagrangian for the motion of a particle in the Newtonian universe, given by Eq.~(1) in Ref. \cite{JMathPhys.56.092501}. Thus, from Eqs.~(\ref{eq:result_1_Newtonian_universe}) and (\ref{eq:action_momentum_Newtonian_universe}), we get
\begin{equation}
F^{2}=-\frac{C_{0}}{\hbar\dot{R}}\ ,
\label{eq:result_2_Newtonian_universe}
\end{equation}
which explicitly shows the dependence of the function $F$ with the scale factor.
%
%
\section{The expansion of the universe from the quantum Newtonian cosmology}\label{Sec.III}
In this section we investigate the classical evolution laws of the universe, which were already obtained using the solutions of the Friedmann equation. However, we want to show that this can also be made from the quantum dynamical interpretation of the Newtonian universe.

Now, consider the wave function of the Schr\"{o}dinger equation in the $R \gg 1$ limit, which implies that $x \gg 1$. The biconfluent Heun functions have the following asympotic behaviors
\begin{equation}
\mbox{HeunB}(\alpha,\beta,\gamma,\delta;x) \sim \left\{
\begin{array}{l}
	x^{\frac{\gamma-2-\alpha}{2}}\sum_{k \geq 0}^{\infty}\frac{a_{k}}{x^{k}}\ ,\\
	\\
	x^{\frac{-\gamma-2-\alpha}{2}}\mbox{e}^{x^{2}+\beta x}\sum_{k \geq 0}^{\infty}\frac{e_{k}}{x^{k}}\ ,
\end{array}
\right.
\label{eq:HeunB_infty_Newtonian_universe}
\end{equation}
where $|\arg x| \leq \frac{\pi}{2}-\epsilon$, $a_{0}=1$, and $e_{0}=1$. Then, the Newtonian wave function can be written as 
\begin{equation}
\psi(R) \sim C_{1}\ (\tau R)^{-\frac{1}{2}+\frac{\gamma}{2}}\ ,
\label{eq:psi_R_HeunB_Newtonian_universe}
\end{equation}
and as a consequence
\begin{equation}
|\psi^{2}(R)| = \frac{C_{1}^{2}}{\tau R} = F^{2}\ ,
\label{eq:psi_R_2_HeunB_Newtonian_universe}
\end{equation}
where we have used the fact that $\gamma$ is an imaginary number. Thus, taking into account Eq.~(\ref{eq:result_2_Newtonian_universe}), we get
\begin{equation}
\frac{\dot{R}}{R}=-\frac{C_{0}\tau}{C_{1}\hbar}\ ,
\label{eq:evolution_Newtonian_universe}
\end{equation}
where $C_{0} < 0$ \cite{PhysLettB.748.361}. Rewriting this formula as
\begin{equation}
\frac{dR}{R}=\frac{|C_{0}|\tau}{C_{1}\hbar}\ dt\ ,
\label{eq:separated_Newtonian_universe}
\end{equation}
its integration give us the following asymptotic behavior for the scale factor
\begin{equation}
R \propto \mbox{e}^{t+t_{0}}\ .
\label{eq:scale_factor_asymptotic}
\end{equation}
The evolution law of the universe from the quantum Newtonian cosmology in the classical limit ($R \gg 1$) is completely consistent with the solution of the Friedmann equation for the vacuum energy predominance, which means that the universe will behave according to the energy contained in the vacuum, whatever the dominant form of energy. In fact, regardless of whether the vacuum energy is due to the cosmological constant or a dark sector, is the component of the vacuum that governs the expansion of the universe when $R$ goes to infinity. Furthermore, our result is independent of the form of energy density and hence it is general than the ones found in the literature \cite{dInverno:1998}.

If we try to fix the constants $C_{0}$ and $C_{1}$ in a such way that the exponential becomes dimensionless, we find
\begin{equation}
C_{1}=\sqrt{\frac{|C_{0}|\mu}{\hbar^{2}\tau}}\ ,
\label{eq:C1}
\end{equation}
where we have used the fact that $E_{n} \sim \hbar^{2}\tau^{2}/\mu$. In this way, we can write
\begin{equation}
R(t)=\mbox{e}^{\frac{\hbar\tau^{2}}{\mu}(t+t_{0})}\ .
\label{eq:full_R}
\end{equation}

Next, consider the wave function of the Schr\"{o}dinger equation in the $R \ll 1$ limit, which implies that $x \ll 1$. In this case, the biconfluent Heun functions have the following asympotic form
\begin{equation}
\mbox{HeunB}(\alpha,\beta,\gamma,\delta;x)=\sum_{s \geq 0}\frac{D_{s}}{(1+\alpha)_{s}}\frac{x^{s}}{s!}\ ,
\label{eq:Biconfluent_Heun_expansion}
\end{equation}
where $D_{0}=1$, and
\begin{eqnarray}
(1+\alpha)_{s} = \frac{\Gamma(s+1+\alpha)}{\Gamma(1+\alpha)}\ .
\label{eq:alpha_s_Biconfluent_Heun_expansion}
\end{eqnarray}
Thus, the Newtonian wave function given by Eq.~(\ref{eq:psi_HeunB_Newtonian_universe}) for the small scale factor $R \ll 1$ can be rewritten as
\begin{equation}
\psi(R) \sim \sqrt{\frac{|C_{0}|\mu}{\hbar^{2}\tau}}\ (\tau R)^{\frac{1+\alpha}{2}}\ ,
\label{eq:psi_R_small_HeunB_Newtonian_universe}
\end{equation}
where we have used the same value for $C_{1}$. Thus, the squared modulus of the wave function is given by
\begin{equation}
\psi^{2}(R) = \frac{|C_{0}|\mu}{\hbar^{2}\tau}\ (\tau R)^{1+\alpha}\ .
\label{eq:psi_R_small_2_HeunB_Newtonian_universe}
\end{equation}
Now, taking into account once more Eq.~(\ref{eq:result_2_Newtonian_universe}), we get
\begin{equation}
R^{1+\alpha}\ dR=\frac{\hbar}{\mu\tau^{\alpha}}\ dt\ .
\label{eq:separated_small_Newtonian_universe}
\end{equation}
Therefore, in this limit, the scale factor has the following asymptotic form
\begin{equation}
R(t) \sim \biggl[\frac{(2+\alpha)\hbar}{\mu\tau^{\alpha}}\biggr]^{\frac{1}{2+\alpha}}t^{\frac{1}{2+\alpha}}\ ,
\label{eq:evolution_small_Newtonian_universe}
\end{equation}
which means that when the universe was very small, its behavior depends on the parameter $\alpha$.
%
%
\section{Conclusions}\label{Sec.IV}
In this work we generalized the previous results for the quantum Newtonian cosmology in the sense that we have now the analytical solutions for each kind of energy density, namely, matter, radiation, vacuum, dark energy and quintessence.

The analysis of the effective potentials for $\Lambda > 0$ and $\Lambda = -|\Lambda|$, taking into account different scenarios, show us that these potentials behave in completely different forms. It is worth noticing that the functional forms of the effective potentials when all kind of energy are considered are similar to the cases when only radiation or vacuum is taken into account, in the appropriate limit when $R \rightarrow 0$ or $R \rightarrow \infty$, respectively. Furthermore, the connection between these regions has the functional form which corresponds to the matter predominance. Otherwise for the other components considered separately, namely, vacuum, dark energy and quintessence, the functional form of the corresponding potentials in the limit $R \rightarrow 0$ are completely different from the one when all kind of energy are present. This means that the components of the sources which correspond to radiation, matter and vacuum determine the behavior of the total effective potential obtained when all sources are considered together.

The Newtonian wave function is given in terms of the biconfluent Heun functions and obeys the appropriate boundary conditions. Thus, the polynomial condition for the biconfluent Heun equation were used to obtain the energy levels related to each value of the parameter $\omega$. As to the energy, in all cases there is a common factor which depends on the vacuum energy as well as on the cosmological constant. The other factor contains a term which is similar to the one corresponding to the spherical oscillator for $\omega = 0$ and $\omega = -1$, and an additional term which depends on the radiation density, for $\omega = 1/3$; on the dark energy density, for $\omega = -1/3$; and on the amount of quintessence, for $\omega = -2/3$.

The dynamical interpretation gives the behavior of the scale factor at the end of the expansion, which is in accordance with the correspondence principle applied to quantum cosmology, namely, we recover the classical limit. On the other hand, when the universe was very small, the quantum effects on the scale factor are given in terms of the parameter $\alpha$, which depends on the form of the energy density.

One conclusion of the Section~\ref{Sec.III} is that the evolution of the universe described by quantum Newtonian cosmology, at the classical level, is consistent with the relativistic description when the energy contained in the vacuum dominates over the others forms of energy.

Finally, it is worth commenting that in the Newtonian description of cosmology, which we have used in the present work, the dynamics of the system is not related to the spacetime itself, which in this context is assumed to be flat. Differently from the Newtonian approach, in the relativistic theory, the dynamics of the system is related to the curvature of spacetime, and therefore, it would be interesting to extend our analysis to the  relativistic context. In fact, some preliminary results which concern this extension were already published recently \cite{arXiv:1904.10864v1}, and we expect to publish some others in a near future.

%
%
\begin{acknowledgments}
The authors would like to thank Conselho Nacional de Desenvolvimento Cient\'{i}fico e Tecnol\'{o}gico (CNPq) for partial financial support. H. S. V. is funded through the research Project No. 150640/2018-8. V. B. B. is partially supported through the research Project No. 305835/2016-5. M. S. C. is partially supported through the research Project No. 312251/2015-7.
\end{acknowledgments}
%
%

%
%

\begin{thebibliography}{99}
%
\bibitem{QJMath.5.64} E. A. Milne, Q. J. Math. \textbf{5}, 64 (1934).
\bibitem{QJMath.5.73} W. H. McCrea and E. A. Milne, Q. J. Math. \textbf{5}, 73 (1934).
%
\bibitem{ProcRSocLondA.206.562} W. H. McCrea, Proc. R. Soc. Lond. A \textbf{206}, 562 (1951).
\bibitem{ProcRSocLondA.149.384} E. T. Whittaker, Proc. R. Soc. Lond. A \textbf{149}, 384 (1935).
%
\bibitem{PhysRev.160.1113} B. S. DeWitt, Phys. Rev. \textbf{160}, 1113 (1967).
\bibitem{Wheeler:1968} J. A. Wheeler, \textit{Batelle Rencontres}, edited by B. S. DeWitt \textit{et al}. (Benjamin, New York, 1968), pp.~242--307.
%
\bibitem{PhysLettB.117.25} A. Vilenkin, Phys. Lett. B \textbf{117}, 25 (1982).
\bibitem{PhysRevD.27.2848} A. Vilenkin, Phys. Rev. D \textbf{27}, 2848 (1983).
\bibitem{PhysRevD.28.2960} J. B. Hartle and S. W. Hawking, Phys. Rev. D \textbf{28}, 2960 (1983).
\bibitem{LettNuovoCimento.39.401} A. D. Linde, Lett. Nuovo Cimento \textbf{39}, 401 (1984).
\bibitem{PhysRevD.30.509} A. Vilenkin, Phys. Rev. D \textbf{30}, 509 (1984).
\bibitem{NuclPhysB.239.257} S. W. Hawking, Nucl. Phys. B \textbf{239}, 257 (1984).
\bibitem{NuclPhysB.252.141} A. Vilenkin, Nucl. Phys. B \textbf{252}, 141 (1985).
\bibitem{PhysRevD.32.2511} A. Vilenkin, Phys. Rev. D \textbf{32}, 2511 (1985).
\bibitem{NuclPhysB.264.185} S. W. Hawking and D. N. Page, Nucl. Phys. B \textbf{264}, 185 (1986).
\bibitem{PhysRevLett.57.2244} A. Ashtekar, Phys. Rev. Lett. \textbf{57}, 2244 (1986).
\bibitem{PhysLettA.236.10} S. P. Kim, Phys. Lett. A \textbf{236}, 11 (1997).
\bibitem{PhysRevD.86.063504} N. Pinto-Neto, F. T. Falciano, R. Pereira and E. S. Santini, Phys. Rev. D \textbf{86}, 063504 (2012).
\bibitem{ClassQuantumGrav.30.143001} N. Pinto-Neto and J. C. Fabris, Classical Quantum Gravity \textbf{30}, 143001 (2013).
%
\bibitem{AIPConfProc.743.286} D. Z. Freedman, M. Schnabl and G. W. Gibbons, AIP Conf. Proc. \textbf{743}, 286 (2004).
\bibitem{arXiv:0504072} J. M. Romero and A. Zamora, arXiv:0504072 \textbf{[gr-qc]} (2005).
\bibitem{ProcRSoc.A.463.503} B. Bramson, Proc. R. Soc. A \textbf{463}, 503 (2007).
\bibitem{IntJTheorPhys.47.455} H. T. Elze, Int. J. Theor. Phys. \textbf{47}, 455 (2008).
\bibitem{ISRNMathPhys.2013.509316} C. Kiefer, ISRN Math. Phys. \textbf{2013}, 509316 (2013).
%
\bibitem{EurPhysJC.76.543} P. Bargue\~no, S. Bravo Medina, M. Nowakowski and D. Batic, Eur. Phys. J. C \textbf{76}, 543 (2016).
%
\bibitem{JMathPhys.56.092501} H. S. Vieira and V. B. Bezerra, J. Math. Phys. \textbf{56}, 092501 (2015).
%
\bibitem{PhysRevD.94.023511} H. S. Vieira and V. B. Bezerra, Phys. Rev. D \textbf{94}, 023511 (2016).
%
\bibitem{Ronveaux:1995} A. Ronveaux, \textit{Heun's Differential Equations}, (Oxford University Press, New York, 1995).
%
\bibitem{AnnPhys.350.14} H. S. Vieira, V. B. Bezerra and C. R. Muniz, Ann. Phys. (NY) \textbf{350}, 14 (2014)
%
\bibitem{PhysLettB.748.361} D. He, D. Gao and Q. Y. Cai, Phys. Lett. B \textbf{748}, 361 (2015).
%
\bibitem{PhysRevD.33.3560} A. Vilenkin, Phys. Rev. D \textbf{33}, 3560 (1986).
%
\bibitem{Bransden:2000} B. H. Bransden and C. J. Joachain, \textit{Quantum mechanics}, (Pearson education, Dorchester, 2000).
%
\bibitem{dInverno:1998} R. d'Inverno, \textit{Introducing Einstein's relativity}, (Oxford University Press, New York, 1998).
%
\bibitem{arXiv:1904.10864v1} H. S. Vieira, V. B. Bezerra, C. R. Muniz and M. S. Cunha, arXiv:1904.10864v1 \textbf{[gr-qc]} (2019).
%
\end{thebibliography}
\end{document}